\documentstyle[psfig,aasms4]{article}
 
\slugcomment{resubmitted to Science,  March 12, 2003} 
 
\begin{document}
 
\title{A Molecular Einstein Ring: Imaging a Starburst Disk Surrounding
a Quasi-Stellar Object}

\author{C. L. Carilli}
\affil{National Radio Astronomy Observatory, P.O. Box O, Socorro, NM,
87801, USA \\
ccarilli@nrao.edu}
\author{G.F. Lewis}
\affil{School of Physics, University of Sydney, NSW 2006, Australia}
\author{S.G. Djorgovski \& A. Mahabal}
\affil{Astronomy Department, California Institute of Technology,
Pasadena, CA, 91125, USA} 
\author{P. Cox}
\affil{Institut d'Astrophysique Spatial, Universit\'e de Paris XI,
91405 Orsay, France}  
\author{F. Bertoldi}
\affil{Max-planck Institut f\"ur Radioastronomie, Auf dem H\"ugel
69, Bonn, D-53121, Germany}
\author{A. Omont}
\affil{Institut d'Astrophysique de Paris, CNRS, 98 bis boulevard
Arago, F-75014, Paris, France} 

\vskip 0.4in

{\bf Images of the CO 2--1 line emission, and the radio continuum
 emission, from the redshift 4.12 gravitationally lensed quasi-stellar
 object (QSO) PSS J2322+1944 reveal an Einstein ring with a diameter
 of 1.5$''$.  These observations are modeled as a star forming disk
 surrounding the QSO nucleus with a radius of 2 kpc. The implied
 massive star formation rate is 900 M$_\odot$ year$^{-1}$.  At this
 rate a substantial fraction of the stars in a large elliptical galaxy
 could form on a dynamical time scale of 10$^8$ years. The observation
 of active star formation in the host galaxy of a high-redshift QSO
 supports the hypothesis of coeval formation of supermassive black
 holes and stars in spheroidal galaxies. }

\vfill\eject

Establishing a link between galaxy formation and massive black hole
formation has become paramount for observational astronomy since the
discovery of the correlation between black hole mass and stellar bulge
mass in nearby ($z < 0.1$) galaxies (1). This correlation suggests a
`causal connection between the formation and evolution of the black
hole and the bulge' (2), and has led to the hypothesis of co-eval
formation at high redshift ($z > 2$) of massive black holes and
spheroidal galaxies (3,4,5). Supermassive black holes ($\ge 10^9$
M$_\odot$) at high redshift manifest themselves as optically luminous
quasi-stellar objects (QSOs) powered by mass accretion onto the hole.

Observations of high redshift QSOs have shown that 30$\%$ are luminous
infrared sources, with far infrared (FIR) luminosities $\sim 10^{13}$
L$_\odot$, corresponding to thermal emission from warm dust (6-9).
The key question for these FIR luminous QSOs is: what is the dominant
dust heating mechanism, star formation or the active galactic nucleus
(AGN)?  If the dust is heated by star formation, the star formation
rates must be of order 10$^3$ M$_\odot$ year$^{-1}$, supporting the
idea of co-eval formation of the stars and black holes in these
systems. Molecular (CO) line emission has been detected from a number of
these FIR-luminous high redshift QSOs, implying large reservoirs of
molecular gas ($> 10^{10}$ M$_\odot$), suggesting that star formation
may be inevitable (9).  However, in most cases the FIR luminosity
corresponds to only 10$\%$ of the bolometric luminosity of the system,
and hence the case for co-eval star formation and mass accretion
onto a supermassive black hole at high redshift remains circumstantial.

The QSO PSS 2322+1944 at $z = 4.12$ is among the most IR-luminous high
redshift QSOs, with an apparent FIR luminosity of $3 \times 10^{13}$
L$_\odot$ (11). Optical imaging and spectroscopy shows that 2322+1944
is a double source, with two essentially identical spectrum components
separated by about 1.5$''$ (10), indicating strong
gravitational lensing by an intervening galaxy.  PSS 2322+1944 is also
the brightest known CO line emitting source at $z > 4$, with an implied
molecular gas mass of $2\times10^{11}$ M$_\odot$ (11) (not corrected for lens
magnification).  Non-thermal (synchrotron) radio continuum emission
from 2322+1944 was detected at 1.4 GHz, and the rest frame
radio-through-IR spectral energy distribution of 2322+1944 matches
closely that of the prototype nuclear starburst galaxy M82
(7,11). Overall, the properties of PSS2322+1944 make it the best
candidate for very active star formation in the host galaxy of a high
redshift QSO. In this paper we present high resolution imaging
of the CO and radio continuum emission from PSS 2322+1944
which address this interesting possibility.

We have observed the CO 2--1 emission from the PSS 2322+1944 at a
resolution of 0.6$''$ using the Very Large Array (VLA).  The
velocity-integrated CO emission from PSS 2322+1944 (Fig. 1) forms a
complete ring with a diameter of about 1.5$''$.  We interpret this
structure as an 'Einstein ring' resulting from strong gravitational
lensing.  The individual velocity channel images (Fig. 2) show that
the CO emission shifts position with velocity, with the peak surface
brightness moving by about 1$''$ from the southeast at low velocity,
to the north at high velocities.  We have also reanalyzed the 1.4 GHz
VLA radio continuum observations of PSS 2322+1944 of (7) by using
different weighting of the visibility data in order to optimize
spatial resolution, as opposed to the previous analysis which
optimized sensitivity at the expense of resolution (Fig. 3). At this
improved resolution (1.1$''$), it is clear that the 1.4 GHz emission
forms a structure similar in size and shape to that seen in the CO
emission.

The unknown characteristics of the lensing galaxy toward 2322+1944 do
not warrant a detailed gravitational lensing inversion.  Einstein
rings, however, are generic features of gravitational lens models (12,
13) and hence we adopt a typical gravitational lens model in our
analysis. In particular, we adopt a lens potential of the form found
for the strongly lensed radio galaxy MG 1131+0456 by (14), including a
mass distribution with an ellipticity of 0.15 at an angle of 36$^o$,
and a central velocity dispersion of 230 km s$^{-1}$.  By examining
various source configurations, and by considering the individual
channel maps, the observed ring structure is reconstructed (Fig. 4).

In this model the optical QSO is located between the inner and outer
caustics for the lens (Figure 4) and is imaged into two point sources
with a total magnification of $\sim3.5$.  The observed CO Einstein
ring can be reproduced by a molecular gas distribution corresponding
to an inclined disk surrounding the optical QSO.  Such disks are
characteristic of the molecular gas and dust distribution in
IR-luminous galaxies seen at low redshift (15).  To form the observed
Einstein ring structure, the CO emission must extend south of the QSO
by at least $\sim 0.3''$ in the source plane, corresponding to
$\sim2.2$kpc assuming a concordance cosmology ($H_o$ = 65 km s$^{-1}$
Mpc$^{-1}$, $\Omega_M = 0.3$ and $\Omega_\Lambda = 0.7$), crossing a
significant fraction of the inner caustic (16).  The scale of the
caustic structure relative to the critical lines is typical of lens
models and hence the lensing mass distribution would have to be
pathologically different to other lens systems to significantly modify
this conclusion.  The total magnification factor for the CO emission
is $\sim2.5$, but this number is dependent upon the thickness of the
CO disk (or inclination angle), as thinner disks will be more strongly
magnified.  The CO component on the northern side of the QSO (away
from the inner caustic) crosses the outer caustic.  This outer
component forms two features in the image plane.  The magnification
for this northern component is less than that for the southern
component by a factor of 2.5 in this model.  This model reproduces the
basic features of the observed structures, including the relative
locations and separations of the double optical QSO and the CO
Einstein ring, and the positions and relative strengths of the three
peaks in the CO distribution (Fig. 1).

The CO emission presents different structure at different velocities
(Fig. 2) consistent with the source straddling the central diamond
caustic. We have considered the source plane distribution of CO as a
function of velocity using the lens model above, and find that the
results are consistent with a disk with a velocity gradient of 115 km
s$^{-1}$ kpc$^{-1}$, implying a mass enclosed within 2.2 kpc of $\rm
3\times 10^{10}\times sin^{-2}$$i$ $M_\odot$, where $i$ is the disk
inclination angle with respect to the plane of the sky.  The center of
the lensing galaxy should lie close to that of the CO ring, roughly
0.5 arcseconds north of the southern QSO image.  Assuming a (likely)
lens redshift of $z \sim 1$, its inferred velocity dispersion is $\sim
230$km s$^{-1}$, suggesting gravitational lensing by a single galaxy.
Hence PSS 2322+1944 represents a prime candidate for a 'Golden Lens'
(17) from which Hubble's constant can be determined via the
measurement of the time delay between the quasar images.

An additional critical constraint on the physical interpretation of
this system is that the observed radio continuum morphology is similar
to that of the CO emission, implying similar source plane
distributions.  While the CO and radio continuum emission form similar
rings in the image plane, the radio continuum peak surface brightness
is shifted by $0.37''\pm0.3''$ from the CO peak, where the positional
error is based on the signal-to-noise ratio of the peak in the 1.4 GHz
image. Hence the position offset could be due to noise, or it
could reflect real source-plane differences on sub-kpc scales between
the radio continuum and CO emission.

The ability of this simplistic modeling approach
to recreate the observed CO configuration highlights the generic
nature of Einstein rings in elliptical gravitational lenses.  Most
importantly, our physical interpretation of this system below
relies on the two most robust aspects of these observations and
modeling: (i) the relative extent of the CO source with respect to the
optical QSO in the source plane, and (ii) the rough co-spatiality of
the CO and radio continuum emission.

The telescopic effect of strong lensing provides the unique
opportunity to study the sub-kpc scale structure in the parent galaxy
of this system, and these results have direct relevance to the
critical question of whether the dust in high redshift, FIR-luminous
QSOs is heated directly by the AGN or by star formation.  For PSS
2322+1944 the relative image plane distribution of the optical and CO
emission implies that the strongly magnified CO emitting regions must
be spatially separated from the QSO by about 2 kpc.  The data are
consistent with a model in which the molecular gas is in an inclined
disk surrounding the QSO.  Although high resolution imaging of the
thermal dust continuum emission from PSS 2322+1944 has not been
performed, it is reasonable to assume that the molecular gas and dust
are roughly co-spatial (18).

Dust heating on kpc-scales has long been consider a problem for
point-source (ie.  AGN) heating models, especially for disk-like dust
distributions (15). The primary difficulty is self-shielding and the
(lack of) illumination of an extended disk by an obscured central
source.  Possible solutions to this problem include warped disks (19),
and heating by dust-penetrating soft Xrays (20).  

For PSS 2322+1944 we have also found that the radio continuum emission
is roughly co-spatial with the molecular gas, and not with the optical
QSO.  This morphology is not what is expected for radio emission from
the AGN itself. Non-thermal radio emission directly associated with
supermassive black holes invariable takes the form of a core-jet radio
source, with the dominant radio emission coming from high energy
electrons accelerated in strong shocks in highly collimated
relativistic outflows from the supermassive black hole (21).  In this
case the observed radio morphology would more closely resemble that of
the optical QSO, ie. a double radio source coincident with the optical
QSOs, and perhaps with extensions corresponding to a jet.

The fact that the radio continuum emitting material is co-spatial with
the  molecular gas,  and  presumably  the dust,  in  PSS 2322+1944  is
exactly what  is expected  for a star  forming galaxy.   The molecular
gas, dust, and radio continuum  emission are always well correlated on
kpc scales  in star forming  galaxies (18), originating in  regions of
active star formation. Given the requisite large gas mass to fuel star
formation, and the similarity of  the radio-to-IR SED of PSS 2322+1944
to that  of a  star forming galaxy,  we conclude  that the QSO  in PSS
2322+1944 is surrounded by a starburst disk on a scale of 2 kpc.
These observations provide the most direct evidence to date of active
star formation in the host galaxy of a luminous, high redshift QSO.
The implied massive star formation rate is $\sim 900$ M$_\odot$
year$^{-1}$ based on the thermal IR luminosity.  At this rate a
substantial fraction of the stars in a large elliptical galaxy could
form on a dynamical timescale of $\le 10^8$ years.

\clearpage
\newpage

\centerline{\bf Figure Captions}

F{\scriptsize IG}. {\bf 1}.--- The contour image of the total CO 2--1
emission (rest frame frequency = 230.538 GHz) from PSS 2322+1944 made
from observations with the Very Large Array (VLA). Observations were
made in October and November, 2002 in the C configuration
(max. baseline = 3 km), giving a resolution of 0.6$''$ at the central
observing frequency of 45.035 GHz.  The total observing time was 18
hours. Fast switching phase calibration was employed (150 sec total
calibration cycle time), and phase coherence was monitored using test
cycles on nearby calibrators. At all times the coherence was found to
be better than 85$\%$. The root-mean-square (rms) noise on this image is
0.09 mJy beam$^{-1}$.  The contour levels are a geometric progression
in square root two starting at 0.12 mJy beam$^{-1}$. The Gaussian
restoring CLEAN beam has FWHM = $0.63'' \times 0.55''$ with a major
axis position angle of -30$^\circ$.  The crosses show the positions of
the optical QSOs, and the cross sizes represent the relative
astrometric error (10).

F{\scriptsize IG}. {\bf 2}. --- The spectral channel images of the CO
2--1 emission from PSS 2322+1944.  Seven spectral channels were used
with a channel width of 6.25 MHz (42 km/s), and the center of channel
4 was at 45.035 GHz (z = 4.1191 for CO 2--1).  Low spatial resolution
imaging of CO 1-0 emission from PSS 2322+1944 constrains the line
center to be at $z = 4.1192 \pm 0.0004$, and the line FWHM = 280 km
s$^{-1}$ (8).  Image cubes were synthesized using natural weighting
of the visibilities, and the final image cube was Hanning smoothed
such that adjacent channels are not independent.  The rms noise
per Hanning smoothed channel was 0.13 mJy
beam$^{-1}$. Channel 7 was found to have higher noise than the other
channels, and was removed from the image cube before Hanning
smoothing.  The contour scheme is the same as Figure 1, starting at
0.25 mJy beam$^{-1}$.  The peak surface brightness in the cube is 0.87
mJy, corresponding to an observed brightness temperature of 1.4 K.

F{\scriptsize IG}. {\bf 3}. --- The radio continuum image of PSS
2322+1944 at 1.4 GHz. The rms in this image is 15$\mu$Jy beam$^{-1}$
and the Gaussian restoring beam is circular with FWHM = 1.1$''$.  The
contour scheme is the same as Figure 1, starting at 0.02 mJy
beam$^{-1}$.  The total flux density is 0.12 mJy and the peak surface
brightness is $56$ $\mu$Jy beam$^{-1}$.

F{\scriptsize IG}. {\bf 4}. --- A gravitational lens model for the CO
emission in PSS 2322+1944.  The model is based on the elliptical lens
toward MG 1131+0456 (14).  The left hand panel presents the source
plane distribution, corresponding to the true (ie.  undistorted by
lensing) morphology of the system.  The image plane distribution is
presented in the right hand panel, corresponding to the observed
morphology after being distorted by the gravitational lens.  The
point-like QSO is represented by a five-point star in both panels. The
solid lines are the caustics and critical lines in the source and
image planes, respectively (16). The CO emission is modeled as an
inclined disk ($i \sim 60^o$) around the QSO, with the north and south
parts of the disk being color-coded red and blue, respectively,
corresponding to different velocity regions on opposite sides of the
QSO.

\clearpage
\newpage

\begin{figure}
\psfig{figure=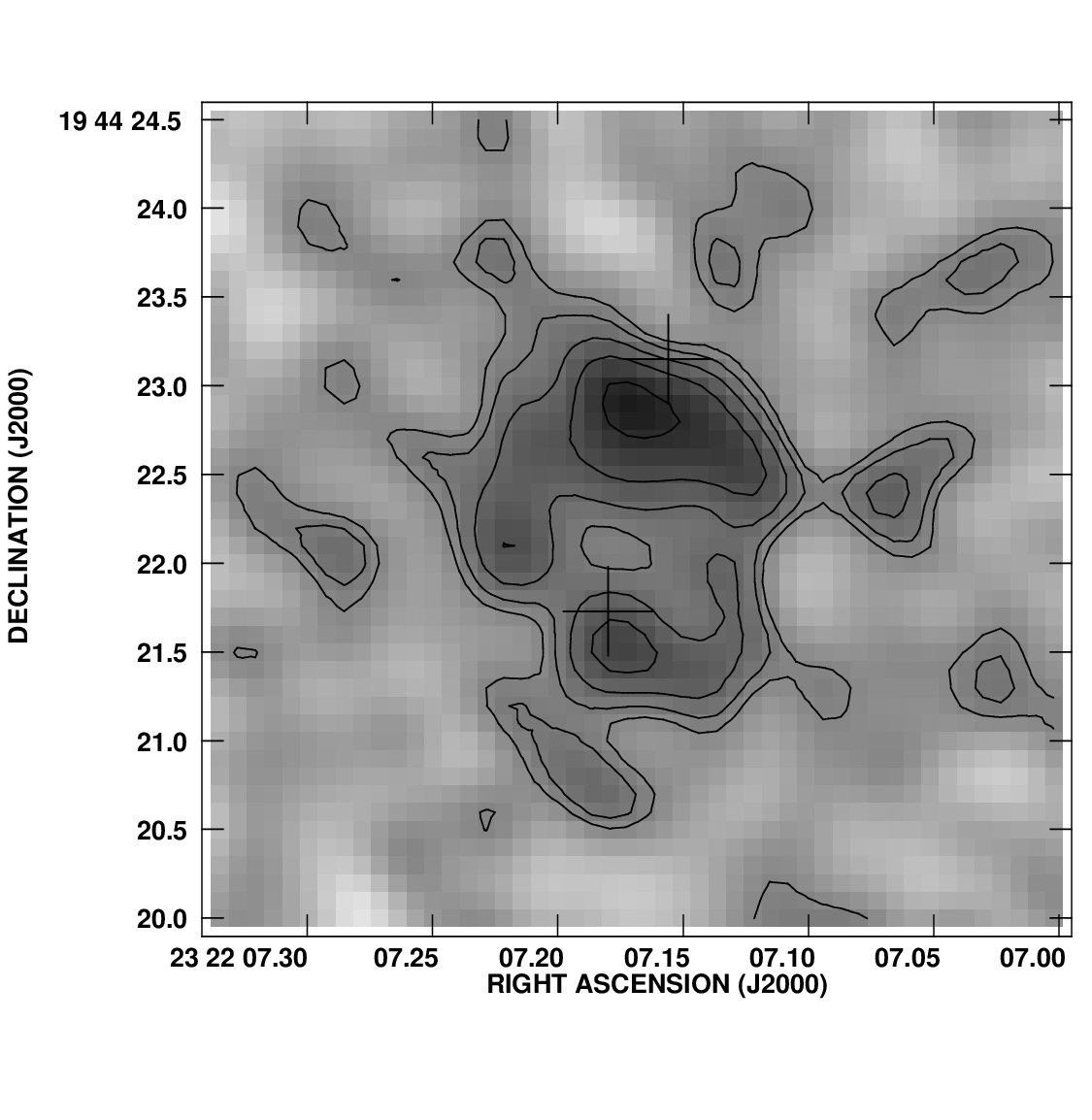,width=6in}
\caption{}
\end{figure}

\clearpage
\newpage

\begin{figure}
\psfig{figure=2322.CUBBEST.PS,width=6in}
\caption{}
\end{figure}

\clearpage
\newpage

\begin{figure}
\psfig{figure=2322.RBEST.PS,width=6in}
\caption{}
\end{figure}

\clearpage
\newpage

\begin{figure}
\psfig{figure=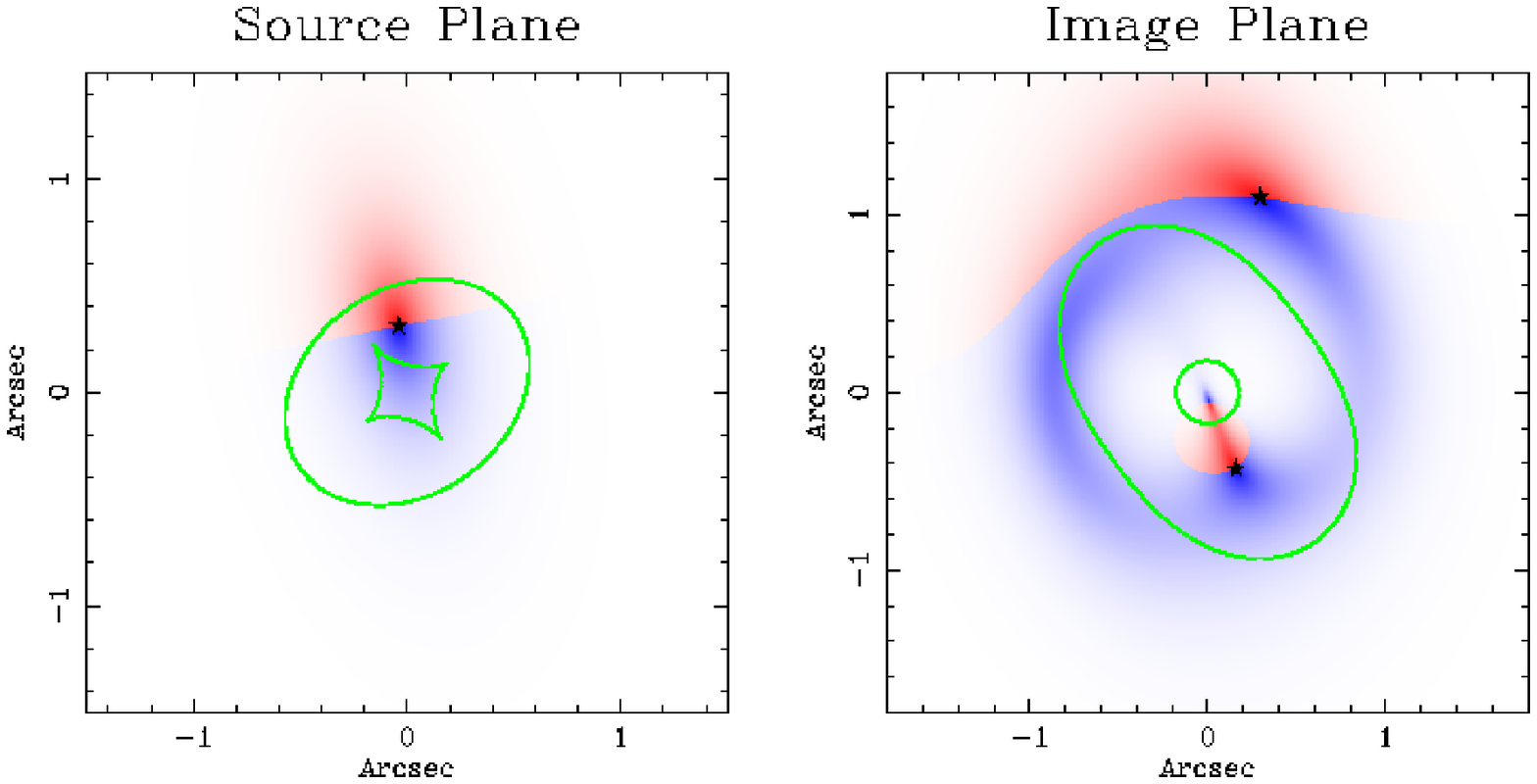,width=7in}
\caption{}
\end{figure}
\clearpage
\newpage

\clearpage
\newpage

\end{document}